\documentclass[%
 reprint,
superscriptaddress,
 amsmath,amssymb,
 aps,
prb,
]{revtex4-2}

\usepackage{graphicx}
\usepackage{float}
\usepackage{dcolumn}
\usepackage{bm}

\usepackage{xcolor}

\usepackage{mathrsfs}
\usepackage{hyperref}
\usepackage{lipsum}
\usepackage{IEEEtrantools}
\usepackage[version=3]{mhchem}
\usepackage{verbatim}
\usepackage{siunitx}

\begin{document}

\preprint{APS/123-QED}

\title{{\em Ab initio} approach for thermodynamic surface phases with full consideration of anharmonic effects -- the example of hydrogen at Si(100)}

\author{Yuanyuan Zhou}
\affiliation{The NOMAD Laboratory at the Fritz Haber Institute of the Max Planck Society, Berlin-Dahlem, Germany}
\author{Chunye Zhu}
\affiliation{The NOMAD Laboratory at the Fritz Haber Institute of the Max Planck Society, Berlin-Dahlem, Germany}
\affiliation{School\ of\ Advanced\ Manufacturing,\ Guangdong\ University\ of\ Technology, Jieyang\ 515200,\ China}
\author{Matthias Scheffler}
\affiliation{The NOMAD Laboratory at the Fritz Haber Institute of the Max Planck Society, Berlin-Dahlem, Germany}
\author{Luca M. Ghiringhelli}

\affiliation{The NOMAD Laboratory at the Fritz Haber Institute of the Max Planck Society, Berlin-Dahlem, Germany}

\date{\today}

\begin{abstract}
A reliable description of surfaces structures in a reactive environment is crucial to understand materials functions. We present a first-principles theory of replica-exchange grand-canonical-ensemble molecular dynamics (REGC-MD) and apply it to evaluate phase equilibria of surfaces in reactive gas-phase environment. We identify the different surface phases and locate phase boundaries including triple and critical points.
The approach is demonstrated by addressing open questions for the Si(100) surface in contact with a hydrogen atmosphere. In the range from 300 to 1\,000~K, we find 25 distinct thermodynamically stable surface phases, for which we also provide microscopic descriptions. Most of the identified phases, including few order-disorder phase transitions, have not yet been observed experimentally. The REGC-MD-derived phase diagram shows significant, qualitative differences to the description by the state-of-the-art ``{\em ab initio} atomistic thermodynamics'' approach.
\end{abstract}

\maketitle


Knowledge of the morphology and structural evolution of materials surfaces in a given reactive atmosphere are prerequisites for understanding mechanism of, e.g.,  heterogeneous-catalysis reactions and crystal growth. In general, reliably tracking of phase equilibria is of technological importance for rational design of surface properties\cite{reuter2016ab}.  \\ 
Studying phase equilibria with first-principles theory is a formidable challenge, especially for highly anharmonic systems, including therein systems with multiple minima separated by shallow barriers.
Pioneering efforts have been performed to estimate the \textit{ab initio} melting line of bulk systems, by direct-coexistence \cite{Correa1204} as well as thermodynamic-integration techniques \cite{PhysRevLett.95.185701, Cheng1110}. Crucially, these techniques require to know the set of relevant phases populating the phase diagram and typically do not yield to the discovery of yet unexpected phases and phase transitions. Similarly, {\em ab initio} atomistic thermodynamics (aiAT) \cite{weinert1986chalcogen, scheffler1, PhysRevB.65.035406, PhysRevLett.111.135501}, the state-of-the-art approach for studying the thermodynamic of surfaces and clusters in reactive environments. It relies on approximations for the vibrational contributions, typically within the harmonic approximation. 

In this letter, we demonstrate a fully {\em ab initio} approach for the construction of surface phase diagrams, including the determination of phase boundaries and triple and critical points. The method entails two steps: data acquisition and data (post) processing. The acquisition of the data is performed by running our recently introduced Replica-Exchange (RE) Grand-Canonical (GC) algorithm \cite{PhysRevB.100.174106}, here coupled with first-principles Born-Oppenhemier molecular dynamics (MD). The REGC approach accounts for all anharmonic contributions without approximations, by performing an unbiased sampling of the configurational and compositional grand-canonical ensemble, i.e., a set of replicas of the studied system are sampled in parallel via MD at different temperatures and chemical potentials (and therefore number of constituent particles) of the gas-phase reacting species. 

\begin{figure*}[!t]
    \centering
    \includegraphics[width=0.95\textwidth]{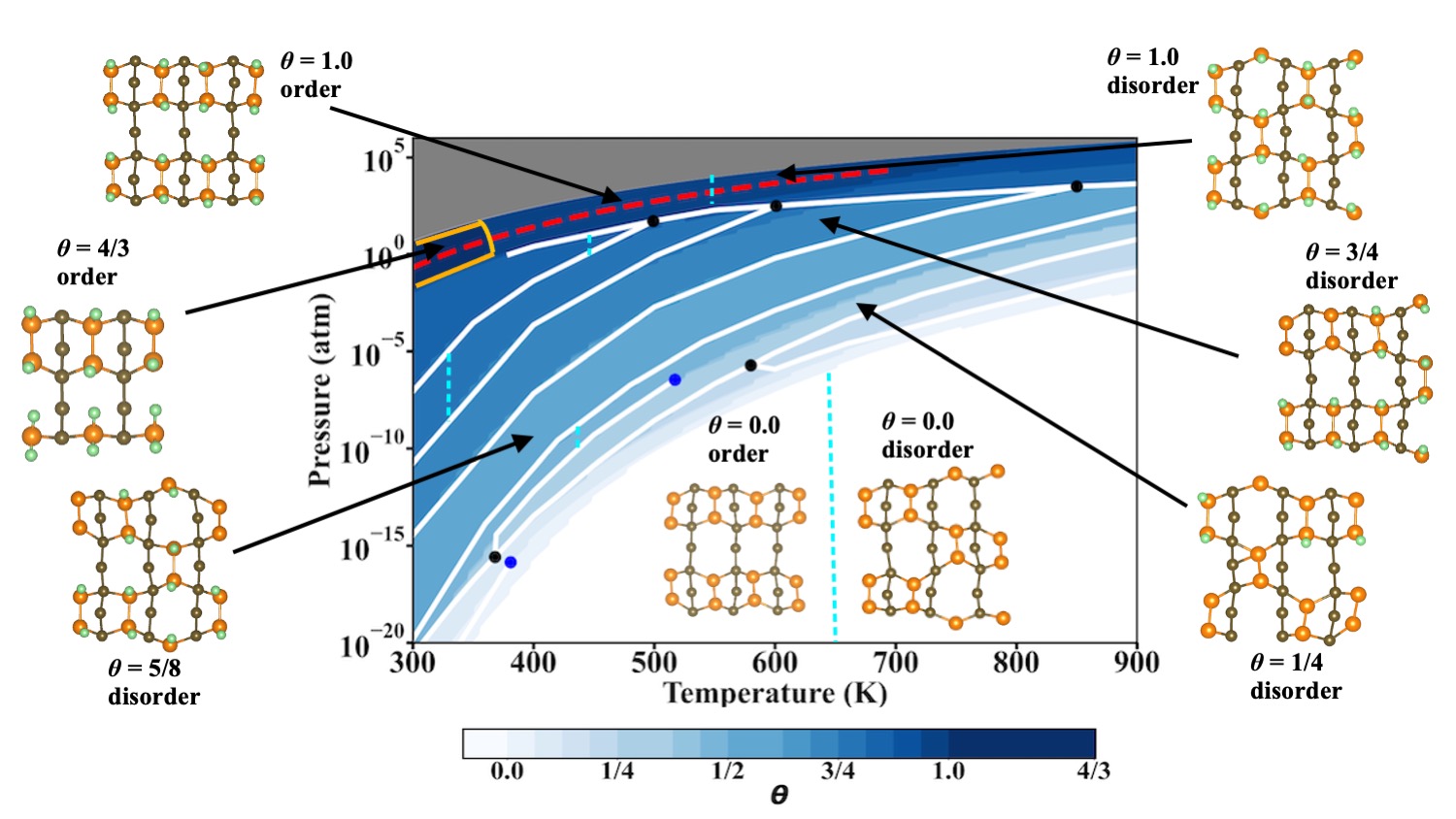}
	\caption{Phase diagram of the Si(100) surface in a D$_2$ gas phase. Each color represents a thermodynamic phase with a certain deuterium coverage. The white area is the stability region of pristine surface. The dark blue area surrounded by the orange frame is the stability region of the phase with maximum coverage of 4/3, sampled in the $3{\times}3$, and its the neighbor phases are sampled in $4{\times}4$ supercell. The white lines indicate the phase boundaries, identified by the analysis of the heat capacity. The critical points are marked as blue dots while the triple points are marked as black dots. The atomic-structure images show the top view of eight representative phases with diverse deuterium coverage $\theta$. The golden spheres are the top silicon atom, the green spheres are deuterium atoms, and the dark spheres are Si atoms in deeper layers. The five cyan dashed lines mark the boundaries of order-disorder phase transitions. The red dashed line indicates $\mu_{\mathrm{D}}=-0.1$ eV, which marks the ($T,p_{\textrm{D}_2}$) path analyzed in Fig. \ref{fig:aiATvsmbar}. The grey area at high pressures corresponds to the region at larger chemical potentials than those sampled in this work.}
    \label{fig:pd1}
\end{figure*}
The REGC approach takes as input only the potential-energy function together with the desired chemical-potential ($\mu$) and $T$ ranges. No prior knowledge about the phase diagram of interest is needed.

Data processing is performed via the multistate-Bennet-acceptance-ratio (MBAR) approach \cite{doi:10.1063/1.1873592}, which is a low-variance estimator of ensemble-averaged thermodynamic observables. The technique is based on the Boltzmann re-weighting and exploits the fact that the configurational density of states is temperature independent, while the probability to observe a given configuration depends on temperature via the Boltzmann factor, $\exp(-\beta U)$ \footnote{Here, we use configurational in the standard statistical-mechanics meaning, i.e., any arrangement of $N$ particles is a configuration and therefore the configurational density of states accounts for the number of particles (specifically in our case, atomic nuclei) arrangements that have an energy (or generalized energy in the case of the grand-canonical ensemble), between $E$ and $E+dE$}, where $\beta = 1/k_\textrm{B}T$ is the inverse temperature and $U$ is the GC potential function of the system, i.e., $U=E- \sum_i \mu_i N_i$. $E$ is the potential energy of a given configuration, $\mu_i$ the chemical potential of each species exchanged with the reservoir, and $N_i$ the number of particles of that species. 
Importantly, all the observables that are processed via MBAR can be identified {\em a posteriori}, i.e., after the data acquisition is completed. MBAR post-processes the data sampled at all temperatures and chemical potential from the REGC-MD run and estimates ensemble values of the desired observables at any given $T$ and $\mu$, not necessarily among the sampled ones.

In order to explain the insight that can be revealed by a REGC {\em ab initio} study, we studied the silicon (100) surface in a hydrogen atmosphere. The chemistry of hydrogen on silicon surfaces has important applications, such as the passivation of surfaces, etching, and CVD growth. Furthermore, the dissociative adsorption of molecular hydrogen on the Si(100) surface has become a paradigm in the study of adsorption systems.\cite{PhysRevLett.74.952, PhysRevLett.81.5596, PhysRevLett.89.166102} 
Three distinct phases have been experimentally observed\cite{neergaard1995surface}: ($i$) a $2{\times}1$ monohydride phase at 600 K (at coverage $\theta=1$, see, e.g., structure at the top left of Fig. \ref{fig:pd1}), where the dimers of the reconstructed pristine Si(100) surface are preserved, ($ii$) a $1{\times}1$ dihydride phase below 300 K where the dimer bonds are broken\cite{BOLAND199217}, and ($iii$) at around 400 K, a $3{\times}1$ phase \cite{PhysRevLett.53.282}, interpreted as alternating rows on monohydrides and dihydrides ($\theta=4/3$, see structure left center in Fig. \ref{fig:pd1}).
Despite the extensive observations under ultra-high vacuum, there is still a lack of systematic measurements to address directly the thermodynamically stable hydrogen-terminated Si(100) structures, when the surface is in equilibrium with an atmosphere of molecular hydrogen at a given temperature and pressure of the reacting atmosphere.

The REGC methodology introduced in Ref. \onlinecite{PhysRevB.100.174106} is here extended by introducing the evaluation of the constant-volume heat capacity and by sampling and jointly post-processing different (here, two) simulations cells, which  allows to overcome a crucial limitation of the GC approach, formally defined only for a constant volume ensemble. 
The evaluation of the heat capacity as function of temperature and pressure of the reacting gas enables us to identify phase-transition lines (narrow stripes in the $(T,p)$ space where the function $C_V(T,p)$ shows ridges) as well as triple and critical points. This concept represents an important advancement compared to 
aiAT, which considers only the differences in free energy between the different phases and locates boundaries where such difference is zero. As we will show below, there are cases where the surface restructures (e.g., changes its coverage) as function of $(T,p)$, but there is no associated peak in $C_V$, and the transition is therefore smooth. 
$C_V$ is calculated via its statistical-mechanics definition:
\begin{equation}\label{eq:specifcheatLJ}
C_{V,\; (T, \;p)} = \frac{\langle E^2\rangle_{(T,\; p)}-\langle E\rangle^2_{(T, \;p)}}{k_BT^2},  
\end{equation}
where $E$ is the DFT total energy (i.e., all potential-energy contributions and kinetic energy) of the system. The ensemble averages of $E$ and $E^2$ are evaluated at each thermodynamical state point ($T$, $p$) of interest, where $p$ is the pressure of the gas in the grand-canonical reservoir, compatible with the sampled $T$ and $\mu$ (see below for more details). This definition is equivalent to the thermodynamic definition $C_V = \left( \partial E / \partial T \right)_V$, as $E$ is $- \partial \ln Z / \partial \beta$, with $Z$ the configurational partition function.  

In MBAR, the ensemble average of any observable $A$ that is function of the configuration $\bm{R}_n$ of the system at a given state point ($\mu$, $\beta$) is statistically evaluated as:
\begin{equation}\label{eq:aveljE}
\langle A \rangle_{\mu,\beta}  =  \sum_{n=1}^{\Omega} \frac{A(\bm{R}_n) \, c_{\mu,\beta}^{-1} \, q(\bm{R}_n;\mu,\beta)}{\sum_{l,m} \Omega_{l,m} c_{\mu_m,\beta_l}^{-1}q(\bm{R}_{l,m};\mu_m,\beta_l)} 
\end{equation}
where $\Omega$ is the total number of samples in all replicas, $\Omega_{l,m}$ is the number of samples in each sampled state point $(m,l)$, where $m$ and $l$ are the index number of the selected chemical potentials and temperatures, respectively. $q(\bm{R}_n;\mu,\beta) = \exp \left[ - U(\bm{R}_n;\mu,\beta) \right]$ is the grand-canonical density function and $c(\mu,\beta)$ is the \textit{partition function} estimated by MBAR \cite{PhysRevB.100.174106, doi:10.1063/1.2978177}. The expectation values of $\langle E \rangle_{\mu,\beta}$ and $\langle E^2 \rangle_{\mu,\beta}$ are evaluated via Eq. \ref{eq:aveljE} where $A(\bm{R}_n)$ is $E(\bm{R}_n)$ and $E^2(\bm{R}_n)$, respectively. 

The first step for the creation of the phase diagram is to evaluate via MBAR the relative free energy as function of $T$ and $\mu$ of all phases, identified within the REGC-MD sampling, as detailed in the SI \cite{supplmentmater} and Ref. \onlinecite{PhysRevB.100.174106}. Here, we limit ourselves the case of only one reactive species exchanged with the reservoir. The different phases are identified by means of structural parameters, e.g., the coverage and the reconstruction/adsorption patterns. The $(T,\mu)$ states are mapped into the more intuitive $(T,p)$ states by choosing the reactive species to be an ideal gas in the reservoir, i.e., $\beta \mu = \ln \Lambda^3 + \ln (\beta p) $, where $\Lambda$ is the thermal wavelength of the species with mass $m$: $\Lambda = h/\sqrt{2 \pi m k T}$.\\ Phase boundaries (white lines in Fig. 1) are estimated by connecting the $(T, \;p)$ state points where the curve $C_V(p)$ at constant $T$, or, symmetrically, $C_V(T)$ at constant $p$, shows a peak \cite{11303_11723}. A critical point is a thermodynamic condition where the free-energy barrier between phases is zero. Here, it is indicated by the $C_V$ peak becoming shallower with increasing $T$ and $p$ then disappearing at ($T_c,\:p_c)$ (critical temperature and pressure). The triple points are the intersections of three phase boundaries. Practical details on how $C_V$ peaks, triple, and critical points are identified are given in the SI \cite{supplmentmater}.\\ 
In our work, the calculated heat capacity maintains a finite value because: ($i$) the simulated system has finite size; ($ii$) the $C_V$ is calculated by integrating over a (small) $\delta p \cdot \delta T$ area where singular point divergences are smeared.

In all our simulations, we use deuterium instead of hydrogen. This is a common approach to lower the vibrational frequencies and thus allow for a longer time-step in the MD runs. For classical particles, this does not change the phase diagram as the configurational partition function depends only on the shape of the potential energy.  Furthermore, we note that similar results are observed in experiments after exposure of silicon surfaces (including the (100) surface) to both H and D\cite{PhysRevLett.53.282, PhysRevLett.54.1055}. This also suggests that the difference in quantum-nuclear effects given by the different isotopes' masses are negligible in the range of temperatures considered in this work. 
We define {\em surface region} as the region of thickness 3.0 \AA~ along the surface normal, starting from average $z$ coordinate of the topmost Si atoms of the surface. The simulation time per replica was 60 ps (3\,000 REGC steps) for Si(100)-($4{\times}4$) (see below for the discussion on the supercell size), resulting in a combined simulation time of around 6 ns. 
All Born-Oppenheimer \textit{ab initio} molecular dynamics (AIMD) trajectories are performed in a canonical ($NVT$) ensemble running for 0.02 ps  and using  a  1-fs  time  step. The  stochastic velocity rescaling thermostat \cite{doi:10.1063/1.2408420} was used to sample the $NVT$ ensemble with $\tau$ parameter of 20 fs. 
All DFT calculations are performed with the all-electron, full-potential electronic-structure package FHI-aims\cite{BLUM20092175}. The Perdew-Burke-Ernzerhof (PBE) \cite{PhysRevLett.78.1396} exchange-correlation functional is used with a tail correction for van der Waals interactions (Tkatchenko-Scheffler scheme) \cite{PhysRevLett.102.073005}. Benchmark studies of model systems and the numerical settings are given in the SI\cite{supplmentmater}. 

For the evaluation of $C_V$ (Eq. \ref{eq:specifcheatLJ}), $\langle E \rangle_{T,\mu}$ and $\langle E^2 \rangle_{T,\mu}$ are calculated at ($T$,$\mu$) with $T$ ranging from  300~K to 1\,000~K (with a spacing of 1~K) and $\mu$ ranging from -0.08 eV to -0.06 eV (with a 0.01 eV spacing). 

The convergence of the sampling was monitored by looking, in each simulated ($T_i,\mu_j$) state, at the average of the distribution of the number of deuterium atoms ($\langle N_\mathrm{D} \rangle$) as well as its variance and verified that these quantities remained constant along the last 80\% of generated data along the MD trajectories, which corresponds to the interval used for the MBAR analysis.

The REGC method is based on the grand-canonical ensemble and thus assumes a fixed volume. To overcome this limitation, we performed two REGC simulations in two different supercells with different volumes. Since MBAR builds the statistics by using the total energy of the samples systems, one needs to identify one structure that is commensurate with both simulation cells, evaluate the energy of such structure in both cells and then shift all energies by subtracting the energy of the chosen common structure in the respective cell. 
Here, as common reference structure, we chose the unreconstructed ($1{\times}1$) pristine Si(100) (see SI for more details).\\
Fig. \ref{fig:pd1} summarizes our main finding, i.e., the phase diagram of Si(100) in the temperature range between 300 and 1\,000~K and in D$_2$ pressure range between $10^{-20}$ and $10^5$ atm. Eight selected phases are indicated in the insets,  but we identified 25 phases, differing in coverage and/or bond connectivity.
The phases are characterized by three structural descriptors: the number of chemisorbed deuterium atoms $N^c_\mathrm{D}$, the coordination histogram $H_{\mathrm{coord}}$ of top layer Si atoms and the dimerization type. A deuterium atom is regarded as chemisorbed onto the silicon surface when the distance to the closest Si atom is smaller than 2.0 \AA. The value of the cutoff is determined as the first minimum in the D-Si radial distribution function of surface structures sampled by REGC (see SI). $H_{\mathrm{coord}}$ is obtained by constructing a coordination histogram (distribution of Si-atoms coordination number) for each surface configuration, i.e., the number of Si atoms bonded to each Si atom and the number of D atoms bonded to each Si atom. The dimerization type is either {\em order} when all the Si-Si dimers are formed on the same side (see, e.g., top left structure in Fig. \ref{fig:pd1}), or \textit{disorder} (e.g., top right structure). At each ($T$, $p_{\mathrm{D_2}}$), the phases are distinguished by the different values of any of the 3 descriptors (for $H_{\mathrm{coord}}$ any change in any bin of the histogram is regarded as a different phase) and the phase at lowest free energy (according to the MBAR evaluation), is reported on the $T$-$p_{\mathrm{D_2}}$ phase diagram.

An order-disorder transition is identified at constant coverage for $\theta = \left\{ 0, 1/2, 13/16, 7/8, 1 \right\}$. For these coverages, the low-temperature, ordered phases consist of Si-Si dimer bonds arranged in stripes, while in the disordered phases, the dimer bonds break and form dynamically between top-layer Si atoms. The total number of dimer bonds remains on average the same across both transitions but the topology is different. Also at other coverages a order-disorder phase transition can exist, but the mentioned five coverages are all for which such transition is found in the $(p,T)$ region of stability for the given coverage. As mentioned, in the disordered phases, the dimer bonds are dynamic. For instance, the average lifetime of the bonds at 800K is $1.48\pm0.05 $ ps and $1.63\pm 0.06$ ps for $\theta=0$ and $\theta=1$, respectively. The longer bond lifetime for $\theta=1$ tells us that adsorbed D stabilizes the dimer bond, i.e, it makes the dynamics of dimer bonds slower (see SI for details on this statistical analysis). \\
As anticipated, another important feature of the phase diagram is that not all changes in coverage are accompanied by a phase boundary, i.e., some surface-structure transformations are smooth. For instance, between $\theta=7/8$ and $\theta = 15/16$ above $\sim 650$~K or between $\theta=3/8$ and $\theta = 1/2$ above the critical point at $\sim 500$ K. In the phase diagram, one can notice the change in color marking the change in coverage, but without a phase boundary in between. Also all the order-disorder phase transitions we found are not accompanied by a $C_V(T)$ peak and are therefore smooth transitions. \\
We note that, in the present approach, the heat capacity cannot be evaluated between phases sampled in two different simulation. Therefore we are unable to determine whether there is a phase boundary between these phases. In the studied system, this is the case of the $\theta = 4/3$ phase, simulated in the $3\times3$ supercell, as compared to the other phases, all simulated in the $4\times4$ supercell. For this reason, we have marked the boundary with a distinctive orange frame.

\begin{figure}[!h]
    \centering
    \includegraphics[width=1.0\columnwidth]{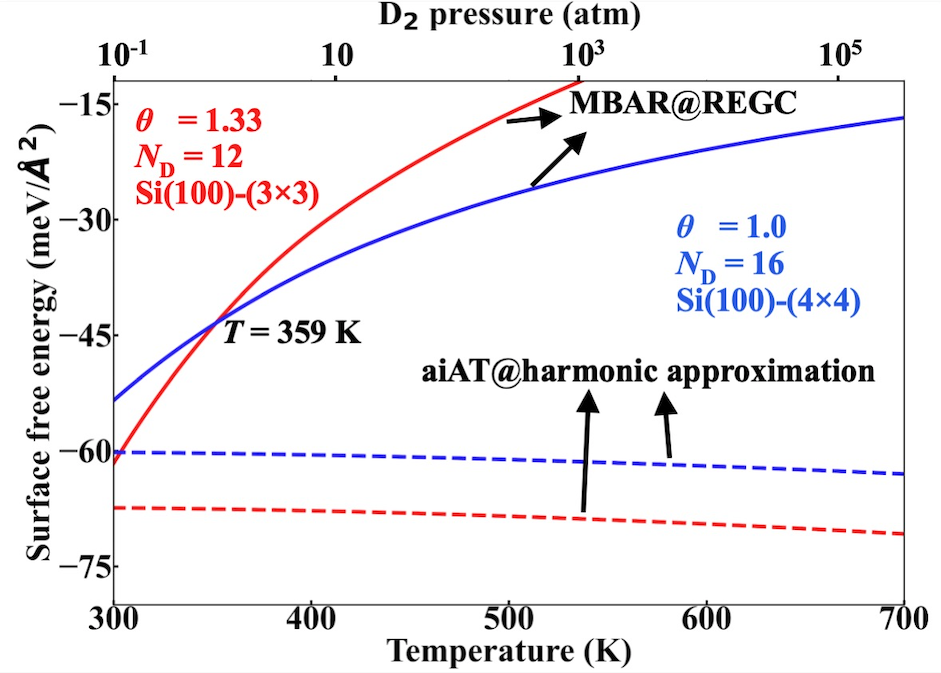}
	\caption{Surface free energies of the Si(100) with different D coverage $\theta = $4/3 (red lines) and $\theta = 1.0$ (blue lines) as a function of temperature at $\mu_{\mathrm{D}}=-0.1$ eV. $\mu_{\mathrm{D}}$ is chosen inside the range where both coverages are stable across a range of relevant $T$. This requirement is met with $\mu_{\mathrm{D}}$ in the range between -0.1 eV and -0.01 eV. The red/blue dash lines are the surface free energy calculated by \textit{ab initio} thermodynamic method at harmonic approximation. The red/blue solid lines are the surface free energy calculated by REGC method. The reference is the unreconstructed bare surface.}
    \label{fig:aiATvsmbar}
\end{figure}

Finally, we address the relative stability of the different surface reconstructions and coverages. We fix our attention at a specific value of $\mu_\textrm{D}=-0.1 $ eV (red dashed line in Fig. \ref{fig:pd1}), chosen to be inside the range of deuterium chemical potentials where both $\theta\!=\!1$ and $\theta\!=\!4/3$ are stable across a wide range of temperatures.
As shown in Fig. \ref{fig:aiATvsmbar}, according to the REGC results, the higher saturation coverage $\theta\!=\!$4/3 is thermodynamically more stable below 359 K than that of $\theta\!=\!1$ and the latter phase becomes thermodynamically more stable when $T>359$ K. The predicted results of REGC is consistent with the experimental observations that $3{\times}1$ LEED pattern is produced at $380 \pm 20$ K and a mild annealing to 600 K yields a very sharp $2{\times}1$ pattern.\cite{PhysRevLett.54.1055, PhysRevB.14.1593, PhysRevB.27.4110}. \\
As a comparison, we show the prediction via {\em ab initio} atomistic thermodynamics (aiAT) for the same two phases, where the vibrational free energy is modeled via the harmonic approximation.
The aiAT results show that two phases coexist in the temperature range from 300 K to 600 K, with only less than 10 meV/\AA$^2$ free energy difference between two phases. The discrepancy between the two methods is ascribed to the anharmonic contributions. These manifest themselves in terms of a complex dynamics of the Si dimers, both in the pristine surface and in the high-coverage surfaces (see discussion of the order parameter dimerization type). Such surface restructuring is completely missed within the harmonic approximation, resulting in aiAT not only failing to account for the phase transition between $\theta = $4/3 and $\theta=1$ phases, but also in erroneously predicting that both coverages become more and more stable with respect to the pristine surface at increasing temperature. In fact, for both $\theta = $ 4/3 and $\theta=1$ phases, the slopes of MBAR@REGC surface free energies as function of $T$ and relative to the pristine surface are positive. 
The dynamical restructuring of Si-Si dimers has more (configurational) freedom in the $4{\times}4$ supercell than that of $3{\times}3$ supercell, thus the $\theta = 1$ phase becomes more stable at high temperature. This is also consistent with the shorter average Si-Si dimer bond lifetime in the pristine compared to the terminated structures. No uniform $1{\times}1$ dihydride phase is found to be stable at any condition, which is consistent with experimental observation.\cite{BOLAND199217} 
 
 In conclusion, we have presented a first-principles theory of Replica-Exchange Grand-Canonical (REGC) molecular dynamics to evaluate and characterize the atomistic structure, composition and geometry, of surfaces in reactive environments, at technologically relevant ($T$, $p$) conditions, including vibrational free energies and all anharmonic effects. The capacity and strength of the approach is demostrated by studying the phase diagram of the Si(100) surface in a deuterium atmosphere. The established ($2{\times}1$) and the controversial ($3{\times}1$) adsorption structures are  found to be thermodynamically stable around $T= 600$~K and 400~K, respectively. 
 Furthermore, a new, dynamic type of Si(100) surface reconstruction is identified at higher temperature, where Si-Si dimer bonds are dynamically restructuring. 
Specifically, three new adsorption patterns are found to be stable at higher temperature and pressure, these are a ($2{\times}2$) and two different ($2{\times}4$) monohydride structures. The approach not only rigorously accounts for the anharmonic vibrational contributions to the free energy but also quantitatively addresses phase boundaries including triple and critical points of a surface at catalytic ($T$, $p$) conditions. Several of the identified phases have not been found experimentally so far.  Examples include low coverage and disorder phases. A possible reason for the lacking experimental results is that the REGC method accesses stable and dynamic surface restructuring in thermodynamic equilibrium with the given $T$ and $p_{\mathrm{D_2}}$ while the experimental studies have been performed in a UHV chamber. Furthermore, it may be experimentally difficult to achieve full thermodynamic equilibrium within the applied measuring times. Our computational REGC approach provides an accurate and robust roadway for fully first-principles predictions of thermodynamic properties relevant for a plethora of important applications, such as heterogeneous catalysis, dopant profiles, surface segregation, and crystal growth.
 
We thank Sergey Levchenko and Haiyuan Wang for useful discussions about the setup of the surface modelling. 
This project has received funding from the European Union's Horizon 2020 research and innovation program (No. 951786: the NOMAD Center of Excellence and No. 740233: TEC1p) and the the  EPSRC  Centre-to  Centre  Project  (Grant  reference: EP/S030468/1). We acknowledge computational resources from the North German Supercomputing Alliance (HLRN). In compliance with the FAIR-data principles, all the REGC-MD trajectories are available at 
https://datashare.mpcdf.mpg.de/s/8ys7q3KvWeR2VmQ.

\bibliographystyle{apsrev4-1}
\bibliography{reference}

\end{document}